\documentclass[
    ,final            
  ]
  {aipproc}

\usepackage{amsfonts}

 \usepackage{hyperref}

\providecommand\apj{ApJ}                 
\providecommand\apjs{ApJSupp}                 
\providecommand\cqg{ClassQuantGrav}                       
\providecommand\aap{A\&A}            
\providecommand\mnras{MNRAS}
\providecommand\BASI{BullAstrSocIndia}

\providecommand\prd{Phys. Rev. D.}
\providecommand\physrep{Physics Reports}
\providecommand\nat{Nature}

\newcommand\Omtot{\Omega_{\mathrm{tot}}}
\newcommand\rC{R_{\mathrm{C}}}

\newcommand\hMpc{\mbox{$h^{-1}$ Mpc}}
\newcommand\hGpc{\mbox{$h^{-1}$ Gpc}}

\layoutstyle{8x11single}

\begin{document}

\title{A residual acceleration effect due to an inhomogeneity}


\author{Boudewijn F. Roukema}{
  address={Toru\'n Centre for Astronomy, Nicolaus Copernicus University,
ul. Gagarina 11, 87-100 Toru\'n, Poland}
}

\begin{abstract}
The perturbed Friedmann-Lema\^{\i}tre-Robertson-Walker models allow
many different possibilities for the 3-manifold of the comoving
spatial section of the Universe. It used to be thought that global
properties of the spatial section, including the topology of the
space, have no feedback effect on dynamics. However, an elementary,
weak-limit calculation shows that in the presence of a density
perturbation, a gravitational feedback effect that is algebraically
similar to dark energy does exist. Moreover, the effect differs between different
3-spaces. Among the well-proportioned spaces, the effect disappears
down to third order in several cases, and down to fifth order for the
Poincar\'e dodecahedral space $S^3/I^*$. The Poincar\'e space, that
which also is preferred in many observational analyses, is
better-balanced than the other spaces.
\end{abstract}

\maketitle


\section{Introduction}

What is the nature of the dark energy (for simplicity, let us use the term
``dark energy'' to generically signify either a form of dark energy or
a cosmological constant, i.e. the parameter $\Omega_\Lambda$) that has been detected
empirically to very high significance in observations of the Universe
with several different telescopes and different types of surveys? This
is one of the two key questions of this conference. However, the ``detection''
of dark energy would be meaningless without the underlying 
general-relativistic models of the Universe. These models 
are the Friedmann-Lema\^{\i}tre-Robertson-Walker models.

What did de~Sitter \citep{deSitt17}, Friedmann
\citep{Fried23,Fried24}, Lema\^{\i}tre \citep{Lemaitre31ell} and
Robertson \citep{Rob35} state were the two properties of the constant
curvature 3-manifold of spatial sections of the Universe that would need
to be determined by observations? They referred to both curvature and
topology. It is through the link between curvature and the various
density parameters (e.g. the total density $\Omtot$) that we 
infer from observations that
dark energy must (according to the exact-FLRW model) exist. What is
the role of topology? 
We now know that theoretically, there are many different
possibilities for the 3-manifold of the comoving spatial section of
the Universe for any of the three curvatures 
\citep{LaLu95,Lum98,Stark98,LR99,RG04} (for a shorter introduction, 
see \cite{Rouk00BASI}). For many years, it was thought that topology
has no effect on the Friedmann equation of an FLRW model, since the Einstein
equations are, by definition, local, while topology (used in this context to 
mean the $\pi_1$ homotopy group of a 3-manifold) is a global property
of a 3-manifold. Indeed, for a perfectly homogeneous FLRW model, this
argument appears to be correct.

However, this parallel session is about the fact that the Universe is
not perfectly homogeneous. The planet Earth, the Solar System, the
Galaxy, clusters of galaxies and the cosmic web of large-scale
structure exist and constitute violations of perfect homogeneity. Many
of the contributions in this session constitute work that may lead to (or already
has led to) claims that the present ``dark energy'' is an approximation 
error rather than a physical phenomenon. That is, the Einstein equations
should in principle be solved by an inhomgeneous solution rather than by
an homogeneous solution with perturbations added afterwards.

Is there a topological feedback effect on the Friedmann equation in an
almost-FLRW Universe, i.e. in one that is close to homogeneous but
contains perturbations? An elementary, weak-limit
calculation for a flat universe with one simply-connected direction
shows that in the presence of a density perturbation, a
gravitational feedback effect does exist \citep{RBBSJ06}.
Moreover, it is algebraically similar to dark energy.
In more general 3-manifolds,  several further 
interesting characteristics are found \citep{RBBSJ06,RR09}. 

\section{$T^1$: a flat universe with one simply-connected direction}
First consider a flat spatial section with one simply-connected direction,
i.e. with spatial section $S^1 \times \mathbb{R}^2$, informally known as
the three-dimensional 1-torus $T^1$ \citep[Sect.~3.1,][]{RBBSJ06}.
 \begin{figure}
   \includegraphics[width=0.9\textwidth]{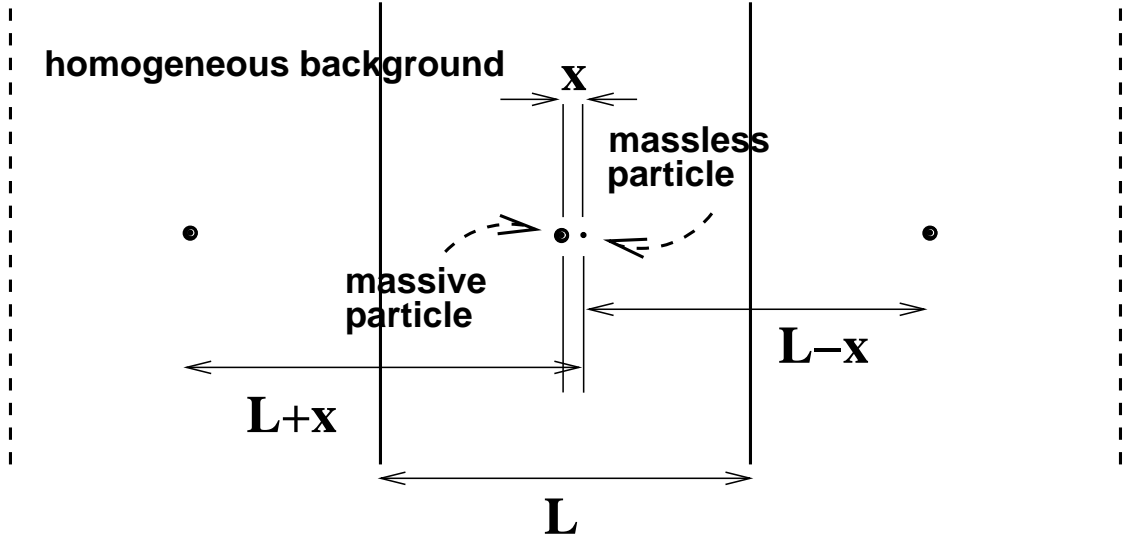}
   \caption{Part of the covering space of a universe with a $T^1$
     3-spatial section, showing three copies of the fundamental domain
     of length $L$. The model is homogeneous except for one massive
     particle (perturbation) as indicated. A massless test particle is
     located at distance $x$ from the massive particle along the short
     fundamental direction of the 3-manifold.
 \label{f-T1}}
 \end{figure}
Figure~\ref{f-T1} shows this schematically. Place a particle of mass $M$,
e.g. a high-mass cluster of galaxies, in an homogeneous
background in a $T^1$ universe of comoving side-length $L$ and consider a
nearby test particle at ``physical'' distance $x$. Assume that the particle horizon
is large enough that gravitational signals from not only the 
nearby, ``original'' instance of the massive object, but also from the two nearest 
distant topological images of the object, have been received.
Approximating this in the weak limit,
it is clear that the acceleration induced by the image of the cluster 
at the right-hand side of the figure will be nearly equal to, but slightly
stronger than, the  acceleration induced by the image of the cluster 
at the left-hand side of the figure. Hence, these two ``topological'' 
accelerations do not quite cancel. The pull to the right is slightly
stronger than that to the left, giving a residual acceleration.
The residual acceleration (in addition
to the acceleration from the copy of the cluster immediately adjacent to the test
particle) is
\begin{eqnarray}
\ddot{x} &=&
            GM \left[ \frac{1}{(L-x)^2} - \frac{1}{(L+x)^2}  \right]
          \nonumber \\
&\approx &  \frac{4GM}{L^2}\;\frac{x}{L}  
\label{e-firstimage-test-object}
\end{eqnarray}
to first order in $x/L \ll 1$, where $G$ is the gravitational constant.

{\em Thus, in the presence of a density perturbation, a gravitational
feedback effect due to global topology can exist.} Spherically symmetric
solutions to the Einstein equations cannot represent this effect, because
the distribution of matter in 
the covering space is not perfectly spherically symmetric. The 
multiple topological images of a single, physical, massive object
prevent any possibility of spherical symmetry.

Algebraically, the residual acceleration effect derived in
Eq.~(\ref{e-firstimage-test-object}) is similar to that of dark
energy: 
\begin{equation}
\ddot{x} \propto x.
\end{equation}
The test particle
sees an effect that opposes its ``normal'' attraction to the 
nearby instance of the massive
particle. The further the test particle is displaced from the (nearby instance of the) massive
particle, the stronger the effect. A fully
relativistic, and necessarily non-spherically-symmetric, derivation of this effect
remains to be carried out, but as derived here in the weak
limit, it is algebraically similar to the effect of dark energy: it is
an acceleration effect. 

The effect is unlikely to explain the presently observed dark energy.
As derived heuristically in Sect.~4.1 of ref~\cite{RBBSJ06}, the
dark energy equation-of-state parameter $w := p/(\rho c^2)$, for pressure $p$,
density $\rho$ and space-time unit conversion constant $c$,
would
be $w \sim - (\chi/L)^3$, where $\chi$ is the comoving separation 
such that $x =  a \chi$ and $a$ is the scale factor.
For a typical distance of an object inside a ``unit'' of large-scale
structure, i.e. a typical distance from a massive cluster of $x \sim 20${\hMpc}, and
a typical value of $L\sim 14${\hGpc} for the 3-manifold favoured from 
Wilkinson Microwave Anisotropy Probe (WMAP) data, either for the
Poincar\'e dodecahedral space 
\citep{LumNat03,Aurich2005a,Aurich2005b,Gundermann2005,Caillerie07,RBSG08,RBG08}
or the 
3-torus $T^3$ 
\citep{WMAPSpergel,Aurich07align,Aurich08a,Aurich08b,Aurich09a},
we have $w \sim -10^{-9}$. Thus, the effect is small at the present.

On the other hand, its role may have been important during early
epochs.  For example, in inflationary scenarios, the pre-inflationary
Universe might have been highly inhomogeneous. The existence
of an effect of topology on the dynamics of an almost-FLRW model shows
that top-down effects, from global to local, can in principle play a
role in physical models of the Universe.

\section{Well-proportioned models}

The simplest way to explain the near-absence of structure in the WMAP
sky maps on scales above $10${\hGpc} (comoving), i.e. the nearly zero
auto-correlation function on these scales (see in particular refs
\citep{Copi07,Copi09}), is that the compactness scale of the Universe
is close to this scale
\citep{LumNat03,Aurich2005a,Aurich2005b,Gundermann2005,Caillerie07,RBSG08,RBG08,WMAPSpergel,Aurich07align,Aurich08a,Aurich08b,Aurich09a}. 
The missing fluctuations interpretation
is a more conservative interpretation of the WMAP cosmological signal
than what is frequently called the ``low quadrupole'' problem or the
``axis of evil'' problem \citep{Copi07,Copi09}. 
The type of 3-manifold that statistically can be expected to 
best match the data is one that is well-proportioned, i.e. that has 
approximately equal fundamental lengths of its fundamental domain in
different directions \citep{WeeksWellProp04}. Indeed, it is the 
regular 3-torus and the Poincar\'e dodecahedral space, both of which 
are well-proportioned, that are at present the favoured models,
although some authors find that an infinite, flat model is 
preferred \citep{KeyCSS06,NJ07}. Among the spherical spaces, there
also exist two other well-proportioned spaces, the 
octahedral space $S^3/T^*$ and the truncated-cube space
$S^3/O^*$ \citep{GausSph01}. What is the nature of the residual
gravitational acceleration effect, as derived above, in these spaces?

\subsection{$T^3$: a flat universe with three simply-connected directions}

In a regular $T^3$ perturbed-FLRW model, i.e. containing a massive particle
as before, consider Fig.~\ref{f-T1}. In addition to the adjacent topological
images to the ``left'' and ``right'', consider images of the massive
particle that are ``below, above,
in front of, and behind'', relative to the 2-plane of the page. All four
of these images are in the 2-plane that exactly bisects the ``left-hand''
and ``right-hand'' images. Hence, the additional gravitational effect of these four images, 
in the weak limit, is to pull the test particle back towards the left, i.e.
it opposes the effect given in Eq.~(\ref{e-firstimage-test-object}).
By how much is the effect weakened? Also, more generally, we need to consider
a test particle offset from the massive particle in an arbitrary direction,
rather than along one of the fundamental axes. In Sect.~3.2 of ref~\citep{RBBSJ06}
and Sect.~3.1 of ref~\citep{RR09}, this more general
calculation is given to first order for slightly irregular $T^3$ models
and to third order for regular $T^3$ models, respectively, using Taylor
expansions in the displacement from the massive particle.

For the regular $T^3$ model,
the opposition between the four ``extra'' images
and the initial images considered in the $T^1$ case leads to the cancellation
of the first-order term in the residual acceleration.
The lowest order terms are the third order terms
\begin{eqnarray}
\ddot{\mathbf{x}}
&=& 
\frac{7GM}{L^2} \; [ \;
  2\epsilon_x^3 - 3\epsilon_x(\epsilon_y^2 + \epsilon_z^2), \;
  2\epsilon_y^3 - 3\epsilon_y(\epsilon_x^2 + \epsilon_z^2), \;
  2\epsilon_z^3 - 3\epsilon_z(\epsilon_x^2 + \epsilon_y^2) \; ],
\label{e-test-object-T3-3rdorder}
\end{eqnarray}
where the massive particle is at $(0,0,0)$ in the covering
space $\mathbb{R}^3$ and the test particle is at
$(x=\epsilon_x L,y=\epsilon_y L,z=\epsilon_z L)$ \{Eq.~(12), \citep{RR09}\}.

In this sense, we can say that a regular $T^3$ model is not only
well-proportioned, but it is also well-balanced: the first-order terms
balance each other perfectly, i.e. they cancel, leaving the
third-order terms to dominate the expression.  This relies on perfect
regularity. 

As shown in Sect.~3.2 of \citep{RBBSJ06}, if the three side-lengths
are unequal, then the linear terms fail to cancel, so that the 
effect is again a linear effect, similar to the $T^1$ case. The effect is
strongest in the shortest direction. It appears that this effect is a
stabilising effect.  Given the heuristic argument in \citep{RBBSJ06},
an irregular $T^3$ model will have slightly anisotropic scale factors
in the different directions in the sense that opposes the
irregularity. Shorter side-lengths will expand a little faster and
longer side-lengths a little slower. Before this result was found,
perfectly regular $T^3$ models were the preferred $T^3$ model
discussed in the literature (e.g.  \citep{Star93,Stevens93}), but
without any physical motivation. The residual gravity effect suggests
a physical motivation for a regular $T^3$ model: it appears to be an equilibrium
point (a stable attractor) in the phase space $(L_a, L_e, L_u, \dot{L_a}, \dot{L_e}, \dot{L_u})$
at which the residual gravity
effect drops from a linear expression (in terms of the dimensionless displacement) to 
a third-order expression. Here,
$L_a, L_e, L_u$ are the three comoving side-lengths and 
$\dot{L_a}, \dot{L_e}, \dot{L_u}$ are their respective derivatives with
respect to proper time (cosmological time).

\subsection{Well-proportioned spherical spaces: the octahedral space $S^3/T^*$,
the truncated-cube space $S^3/O^*$ and the Poincar\'e dodecahedral 
space $S^3/I^*$}

What happens in the spherical well-proportioned spaces?  Calculating
the residual acceleration effect for the spherical well-proportioned
spherical spaces, i.e. the octahedral space $S^3/T^*$, the
truncated-cube space $S^3/O^*$ and the Poincar\'e dodecahedral space
$S^3/I^*$, is most easily done by embedding the covering space $S^3$
in Euclidean 4-space $\mathbb{R}^4$. As explained in Sect.~2.1 of
ref~\citep{RR09}, weak-limit gravity in a spherical covering space is
proportional to $[\rC \sin({r}/{\rC}) ]^{-2}$, where $\rC$ is the
curvature radius, rather than to $r^{-2}$, in order to satisfy Stokes'
theorem.

In Sect.~3.3 of ref~\citep{RR09} it is shown numerically that the octahedral space
and the truncated-cube space are also well-balanced. The residual gravity
effect due to the adjacent images of the massive object balance well enough
that again, the linear term disappears and the effect is dominated by 
third order terms.

The effect in the Poincar\'e dodecahedral space is even more finely
balanced than for the other well-proportioned spaces. In
ref~\citep{RR09}, this is shown numerically, and also
analytically. Both the linear and the third-order terms cancel.
Writing the displacement as a four-vector $\mathbf{r}$
in $\mathbb{R}^4$, 
necessarily lying in the covering space, i.e.
$\mathbf{r} \in S^3 \subset \mathbb{R}^4$,
the dominant term is the {\em fifth}-order term, which can be written as
a vector in $\mathbb{R}^4$
\begin{eqnarray}
 \ddot{\mathbf{r}} &=&
\frac{
12 \sqrt{2} \left(297 \sqrt{5} + 655\right)
}{125 \sqrt{5-\sqrt{5}}}
\left(\frac{GM}{\rC^2}\right) 
\left(\frac{r}{\rC}\right)^5
 \nonumber \\  && 
\Big\{
\big[70\,y^4
+(42\,\sqrt{5}+70)\,x^2\,y^2
-(14\,\sqrt{5}+70)\,y^2    
+(21\,\sqrt{5}-7)\,x^4
-28\,\sqrt{5}\,x^2       
+7\,\sqrt{5}+5\big] \;x,    \nonumber \\  && 
\big[70\, z^4
+(42\,\sqrt{5}+70)\,y^2\,z^2
-(14\,\sqrt{5}+70)\,z^2
+(21 \,\sqrt{5}-7)\,y^4
-28\,\sqrt{5}\,y^2       
+7\,\sqrt{5}+5\big] \;y,     \nonumber \\  && 
\big[
70\,x^4
+(42\,\sqrt{5}+70) \,x^2\,z^2
-(14\,\sqrt{5}+70)\,x^2
+(21\,\sqrt{5}-7)\,z^4
-28\,\sqrt{5}\,z^2        
+7\,\sqrt{5}+5\big] \;z ,     \nonumber \\  && 
0
\Big\} , 
\label{e-residgrav-Poincare-exact}
\end{eqnarray}
where 
the curvature radius is $\rC$, 
the massive particle is at $(0,0,0,\rC)$, 
the nearby test particle is at
$\mathbf{r} := \rC [\sin(r/\rC) x, \sin(r/\rC) y , \sin(r/\rC) z, \cos(r/\rC)]$, 
and $x^2+y^2+z^2=1$
\{cf. Eq.~(21), \cite{RR09}\}. 
Although it appears in 
Eq.~(\ref{e-residgrav-Poincare-exact}) 
that the residual acceleration 
is exactly tangent to the
observer at $(0,0,0,\rC)$, this is only an artefact of showing only
the dominant (fifth-order) term in the residual acceleration.
The dominant term in the fourth ($w$) direction in the embedding space is a 
sixth-order term, and the residual acceleration does
indeed lie in the tangent 3-plane to the 3-sphere at $\mathbf{r}$, as it
must.

Hence, the Poincar\'e dodecahedral space $S^3/I^*$, the spatial 3-section that is
preferred in many observational analyses, is better balanced than the
other well-proportioned spaces. General 3-manifolds are dominated by 
the linear term, most of the (regular) well-proportioned spaces are dominated by 
the third-order term, and the Poincar\'e space is dominated by the fifth-order
term.

\section{Conclusion}

Throughout the history of cosmology since ancient times, preferences
of the model of space have mostly shifted between finite spaces with a
boundary and infinite unbounded spaces. Riemannian geometry made it
possible to have flat (or hyperbolic) finite 3-spaces of constant
curvature without boundaries. Pseudo-Riemannian geometry extends these
to the FLRW space-time models of the Universe. Infinity is not a real
number, and for the Universe to be a physical object, it would be most
reasonable for the comoving volume, and hence total mass-energy, of
the Universe to be finite.

\begin{table}
\caption{Properties of empirically favoured ($\mathbb{R}^3, T^3, S^3/I^*$) spatial 3-sections and some related 3-spaces  \label{t-summary}}
 $ \begin{array}{c c c c } \hline 
      \mathrm{comoving} 
      \rule[-1.5ex]{0ex}{3.5ex}
      & {\mathrm{finite?}}
      & {\mathrm{missing}}
      & \ddot{x}_{\mathrm{resid}} \propto
      \\ 
      \mbox{3-space} 
      & 
      & {\mathrm{fluctuations?}}
      &
      \\
      & 
      & {\mathrm{(WMAP)}}
      &
      \\ \hline 
      \rule{0ex}{2.5ex}  
      \mbox{infinite flat space} \;\;  { { \mathbb{R}^3 }}
      & {\mathrm{ no} }
      & {\mathrm{ no} }
      & {\mathrm{N/A} }
      \\ 
      \mbox{1-torus} \;\;  {{ T^1 := {S^1}\times \mathbb{R}^2 }}
      & {\mathrm{ no} }
      & {\mathrm{ no} }
      & { (x/L)^1}
      \\ 
      \mbox{3-torus} \;\;  {{ {T}^3 }}
      & {\mathrm{ yes} }
      & {\mathrm{ yes} }
      & { (x/L)^3}
      \\ 
        \mbox{octahedral space} \;\;  {{ S^3/T^* }}
      & {\mathrm{ yes} }
      & {\mathrm{ no} }
      & { (x/\rC)^3}
      \\ 
        \mbox{truncated cube space} \;\;  {{ S^3/O^* }}
      & {\mathrm{ yes} }
      & {\mathrm{ no} }
      & { (x/\rC)^3}
      \\
        \mbox{Poincar\'e dodecahedral space} \;\;  {{ S^3/I^* }}
      &{\mathrm{ yes} }
      & {\mathrm{ yes} }
      & { (x/\rC)^5 }
      \\
      \hline
    \end{array}$
\end{table}

The preferred models of comoving space, given the most recent
cosmological observational catalogues, especially including the WMAP
data, are summarised in Table~\ref{t-summary}. The other models
discussed above are included as well. The displacement is written as 
a scalar $x$ in all three cases for simplicity. Full expressions of
the dominant terms for $T^3$ and $S^3/I^*$ are given in 
Eq.~(\ref{e-test-object-T3-3rdorder}) and
Eq.~(\ref{e-residgrav-Poincare-exact}) respectively.
While $\mathbb{R}^3$ is frequently considered to be the implicit spatial
section of the Concordance model, this is rarely stated explicitly,
since it is not usually meant to be interpreted literally as a global model.
It has the physically undesirable characteristic of giving the Universe
an infinite amount of mass-energy, and has difficulty in
explaining the lack of fluctuations above 10{\hGpc} (in projection, 
above 60 degrees on the sky) in the WMAP cosmic microwave background maps.

Among the finite spaces, the well-proportioned spaces are, in general,
good candidates for explaining the lack of $> 10${\hGpc} fluctuations, but constraints
on the curvature radius $\rC$ make the fundamental domains of 
the octahedral and truncated-cube spaces a little too large compared to 
the surface of last scattering. For this reason, ``no'' is written for 
these spaces in Table~\ref{t-summary}. The best candidates would appear
to be the regular 3-torus $T^3$ and the Poincar\'e dodecahedral space
$S^3/I^*$. The latter appears to constitute an extremum among the class
of possible 3-manifolds, in that the residual gravity effect is
exceptionally well balanced, down to fifth order. 
We could say that ``some spaces are more equal than others''
and that the Poincar\'e space is the space that is ``the most equal''
\citep{Orwell45}.

This characteristic
is a gravitational, geometrical, topological 
property of the Poincar\'e space, provided that an inhomogeneity exists
in the space, and it  
was clearly unknown to the group
that first proposed the Poincar\'e space as the best fit to the WMAP data
\citep{LumNat03}. 

Is this just a coincidence? Or is it possible that gravity, geometry,
topology and
inhomogeneity together determine the most likely space to have been realised
by early universe physical processes, in the way derived above, and that
this most likely space is the one first proposed in 2003 in order
to fit the WMAP data \citep{LumNat03}?



\bibliographystyle{aipproc}   

\begin{thebibliography}{34}
\expandafter\ifx\csname natexlab\endcsname\relax\def\natexlab#1{#1}\fi
\providecommand{\enquote}[1]{``#1''}
\expandafter\ifx\csname url\endcsname\relax
  \def\url#1{\texttt{#1}}\fi
\expandafter\ifx\csname urlprefix\endcsname\relax\def\urlprefix{URL }\fi
\providecommand{\eprint}[2][]{\url{ArXiv:#2}}

 \renewcommand{\eprint}[1]{\href{http://arxiv.org/abs/#1}{{\tt [#1]}}}


\bibitem[{de Sitter}(1917)]{deSitt17}
W.~{de Sitter}, \emph{\mnras} \textbf{78}, 3 (1917).

\bibitem[{Friedmann}(1923)]{Fried23}
A.~{Friedmann}, \emph{{{\sl Mir kak prostranstvo i vremya} (The Universe as
  Space and Time)}}, Leningrad: Academia, 1923.

\bibitem[{Friedmann}(1924)]{Fried24}
A.~{Friedmann}, \emph{Zeitschr. f\"ur Phys.} \textbf{21}, 326 (1924).

\bibitem[{Lema{\^i}tre}(1931)]{Lemaitre31ell}
G.~{Lema{\^i}tre}, \emph{\mnras} \textbf{91}, 490 (1931).

\bibitem[{Robertson}(1935)]{Rob35}
H.~P. {Robertson}, \emph{\apj} \textbf{82}, 284 (1935).

\bibitem[{Lachi\`eze-Rey} and {Luminet}(1995)]{LaLu95}
M.~{Lachi\`eze-Rey}, and J.~{Luminet}, \emph{\physrep} \textbf{254}, 135
  (1995), \eprint{gr-qc/9605010}.

\bibitem[{Luminet}(1998)]{Lum98}
J.-P. {Luminet}, \emph{Acta Cosmologica} \textbf{XXIV-1}, 105 (1998),
  \eprint{gr-qc/9804006}.

\bibitem[{Starkman}(1998)]{Stark98}
G.~D. {Starkman}, \emph{\cqg} \textbf{15}, 2529 (1998).

\bibitem[{Luminet} and {Roukema}(1999)]{LR99}
J.~{Luminet}, and B.~F. {Roukema}, \enquote{{Topology of the Universe: Theory
  and Observation},} in \emph{NATO ASIC Proc. 541: Theoretical and
  Observational Cosmology. Publisher: Dordrecht: Kluwer,}, 1999, p. 117,
  \eprint{astro-ph/9901364}.

\bibitem[{Rebou\c{c}as} and {Gomero}(2004)]{RG04}
M.~J. {Rebou\c{c}as}, and G.~I. {Gomero}, \emph{Braz. J. Phys.} \textbf{34},
  1358 (2004), \eprint{astro-ph/0402324}.

\bibitem[{Roukema}(2000)]{Rouk00BASI}
B.~F. {Roukema}, \emph{\BASI} \textbf{28}, 483 (2000),
  \eprint{astro-ph/0010185}.

\bibitem[{Roukema} et~al.(2007)]{RBBSJ06}
B.~F. {Roukema}, S.~{Bajtlik}, M.~{Biesiada}, A.~{Szaniewska}, and
  H.~{Jurkiewicz}, \emph{\aap} \textbf{463}, 861 (2007),
  \eprint{astro-ph/0602159}.

\bibitem[{Roukema} and {R\'o\.za\'nski}(2009)]{RR09}
B.~F. {Roukema}, and P.~T. {R\'o\.za\'nski}, \emph{\aap} \textbf{502}, 27
  (2009), \eprint{0902.3402}.

\bibitem[{Luminet} et~al.(2003)]{LumNat03}
J.~{Luminet}, J.~R. {Weeks}, A.~{Riazuelo}, R.~{Lehoucq}, and J.~{Uzan},
  \emph{\nat} \textbf{425}, 593 (2003), \eprint{astro-ph/0310253}.

\bibitem[{Aurich} et~al.(2005{\natexlab{a}})]{Aurich2005a}
R.~{Aurich}, S.~{Lustig}, and F.~{Steiner}, \emph{\cqg} \textbf{22}, 3443
  (2005{\natexlab{a}}), \eprint{astro-ph/0504656}.

\bibitem[{Aurich} et~al.(2005{\natexlab{b}})]{Aurich2005b}
R.~{Aurich}, S.~{Lustig}, and F.~{Steiner}, \emph{\cqg} \textbf{22}, 2061
  (2005{\natexlab{b}}), \eprint{astro-ph/0412569}.

\bibitem[{Gundermann}(2005)]{Gundermann2005}
J.~{Gundermann}, \emph{ArXiv e-prints}  (2005), \eprint{astro-ph/0503014}.

\bibitem[{Caillerie} et~al.(2007)]{Caillerie07}
S.~{Caillerie}, M.~{Lachi{\`e}ze-Rey}, J.~. {Luminet}, R.~{Lehoucq},
  A.~{Riazuelo}, and J.~{Weeks}, \emph{\aap} \textbf{476}, 691 (2007),
  \eprint{0705.0217v2}.

\bibitem[{Roukema} et~al.(2008{\natexlab{a}})]{RBSG08}
B.~F. {Roukema}, Z.~{Buli\'nski}, A.~{Szaniewska}, and N.~E. {Gaudin},
  \emph{\aap} \textbf{486}, 55 (2008{\natexlab{a}}), \eprint{0801.0006}.

\bibitem[{Roukema} et~al.(2008{\natexlab{b}})]{RBG08}
B.~F. {Roukema}, Z.~{Buli\'nski}, and N.~E. {Gaudin}, \emph{\aap} \textbf{492},
  673 (2008{\natexlab{b}}), \eprint{0807.4260}.

\bibitem[{Spergel} et~al.(2003)]{WMAPSpergel}
D.~N. {Spergel}, L.~{Verde}, H.~V. {Peiris}, E.~{Komatsu}, M.~R. {Nolta}, C.~L.
  {Bennett}, M.~{Halpern}, G.~{Hinshaw}, N.~{Jarosik}, A.~{Kogut}, M.~{Limon},
  S.~S. {Meyer}, L.~{Page}, G.~S. {Tucker}, J.~L. {Weiland}, E.~{Wollack}, and
  E.~L. {Wright}, \emph{\apjs} \textbf{148}, 175 (2003),
  \eprint{astro-ph/0302209}.

\bibitem[{Aurich} et~al.(2007)]{Aurich07align}
R.~{Aurich}, S.~{Lustig}, F.~{Steiner}, and H.~{Then}, \emph{\cqg} \textbf{24},
  1879 (2007), \eprint{astro-ph/0612308}.

\bibitem[{Aurich}(2008)]{Aurich08a}
R.~{Aurich}, \emph{\cqg} \textbf{25}, 225017 (2008), \eprint{0803.2130}.

\bibitem[{Aurich} et~al.(2008)]{Aurich08b}
R.~{Aurich}, H.~S. {Janzer}, S.~{Lustig}, and F.~{Steiner}, \emph{Classical and
  Quantum Gravity} \textbf{25}, 125006 (2008), \eprint{0708.1420}.

\bibitem[{Aurich} et~al.(2009)]{Aurich09a}
R.~{Aurich}, S.~{Lustig}, and F.~{Steiner}, \emph{ArXiv e-prints}  (2009),
  \eprint{0903.3133}.

\bibitem[{Copi} et~al.(2007)]{Copi07}
C.~J. {Copi}, D.~{Huterer}, D.~J. {Schwarz}, and G.~D. {Starkman}, \emph{\prd}
  \textbf{75}, 023507 (2007), \eprint{astro-ph/0605135}.

\bibitem[{Copi} et~al.(2008)]{Copi09}
C.~J. {Copi}, D.~{Huterer}, D.~J. {Schwarz}, and G.~D. {Starkman}, \emph{ArXiv
  e-prints}  (2008), \eprint{0808.3767}.

\bibitem[{Weeks} et~al.(2004)]{WeeksWellProp04}
J.~{Weeks}, J.-P. {Luminet}, A.~{Riazuelo}, and R.~{Lehoucq}, \emph{\mnras}
  \textbf{352}, 258 (2004), \eprint{astro-ph/0312312}.

\bibitem[{Key} et~al.(2007)]{KeyCSS06}
J.~S. {Key}, N.~J. {Cornish}, D.~N. {Spergel}, and G.~D. {Starkman},
  \emph{\prd} \textbf{75}, 084034 (2007), \eprint{astro-ph/0604616}.

\bibitem[{Niarchou} and {Jaffe}(2007)]{NJ07}
A.~{Niarchou}, and A.~{Jaffe}, \emph{Physical Review Letters} \textbf{99},
  081302 (2007), \eprint{astro-ph/0702436}.

\bibitem[{Gausmann} et~al.(2001)]{GausSph01}
E.~{Gausmann}, R.~{Lehoucq}, J.-P. {Luminet}, J.-P. {Uzan}, and J.~{Weeks},
  \emph{\cqg} \textbf{18}, 5155 (2001), \eprint{gr-qc/0106033}.

\bibitem[{Starobinsky}(1993)]{Star93}
A.~A. {Starobinsky}, \emph{Journal of Experimental and Theoretical Physics
  Letters} \textbf{57}, 622 (1993).

\bibitem[{Stevens} et~al.(1993)]{Stevens93}
D.~{Stevens}, D.~{Scott}, and J.~{Silk}, \emph{Physical Review Letters}
  \textbf{71}, 20 (1993).

\bibitem[{Orwell}(1945)]{Orwell45}
G.~{Orwell}, \emph{{Animal Farm: A Fairy Story}}, London: Secker and Warburg,
  1945.

\end{thebibliography}


\end{document}